\documentclass[12pt,preprint]{aastex}
\newcommand{\etal}{et al.}
\newcommand{\teff}{$T_{\!\mbox{\scriptsize \em eff}}$} 
\shorttitle{Spectroscopy of blue supergiants in NGC 300}
\shortauthors{Bresolin, Gieren, Kudritzki, Pietrzy{\'n}ski \&
Przybilla}

\begin{document}

\title{Spectroscopy of blue supergiants in the spiral galaxy NGC
300\footnotemark} \footnotetext[1]{Based on observations obtained at
the ESO Very Large Telescope}

\author{Fabio Bresolin\footnotemark}
\affil{Institute for Astronomy, 2680 Woodlawn Drive, Honolulu HI
96822}
\email{bresolin@ifa.hawaii.edu}

\footnotetext[2]{On leave from Universit{\"a}ts-Sternwarte M{\"u}nchen}

\author{Wolfgang Gieren}
\affil{Universidad de Concepci{\'o}n, Departamento de Fisica, Casilla 160-C,
Concepci{\'o}n, Chile} \email{wgieren@coma.cfm.udec.cl}
\author{Rolf-Peter Kudritzki}
\affil{Institute for Astronomy, 2680 Woodlawn Drive, Honolulu HI
96822} \email{kud@ifa.hawaii.edu}
\author{Grzegorz Pietrzy{\'n}ski\footnotemark}
\affil{Universidad de Concepci{\'o}n, Departamento de Fisica, Casilla 160-C,
Concepci{\'o}n, Chile}
 \email{pietrzyn@hubble.cfm.udec.cl} 
\footnotetext[3]{Also affiliated to: Warsaw University Observatory, Al.
Ujazdowskie 4, 00-478, Warsaw, Poland}
\and
\author{Norbert Przybilla}
\affil{Universit{\"a}ts-Sternwarte M{\"u}nchen, Scheinerstr. 1,
81679, M{\"u}nchen, Germany} \email{nob@usm.uni-muenchen.de}

\begin{abstract}
We have obtained VLT low-resolution ($\sim5$~\AA) multi-object
spectroscopy in the 4,000-5,000 \AA\/ spectral range of about 70
blue supergiant candidates in the Sculptor Group spiral galaxy
NGC~300, selected from previous wide-field photometry. Of the 62
spectroscopically confirmed blue supergiants, with spectral types
ranging from late-O to F, 57 have types between early-B and mid-A. We
present a detailed spectral catalog containing identification,
magnitudes, colors and spectral types.

We employ synthetic spectra calculated from blue supergiant model
atmospheres for different metallicities to determine metal abundances
for two A0 supergiants of the sample. In agreement with the
expectations, the star closer to the galactic center is found to be
more metal rich than the object at a larger galactocentric
distance. We will employ this technique on the whole supergiant sample
to determine the stellar abundance gradient in the disk of
NGC~300, together with the internal reddening from a comparison
of the observed vs. synthetic colors. This will allow, among other
things, an accurate calibration of the effect of metallicity on the
Cepheid Period-Luminosity relation.

Using the Balmer H$\beta$ line profile we have estimated the mass-loss
rate for one of the brightest A2 supergiants in the sample. Under
additional reasonable assumptions we determined the wind momentum of
the star and compared it to the value expected from the empirical wind
momentum-luminosity relationship (WLR) for A-type supergiants of
Kudritzki et al. (1999). Good agreement is obtained. We will derive
mass-loss rates and wind momenta for all stars in our sample from the
H$\alpha$ line profiles in forthcoming work, and will then thoroughly
test the usefulness of the WLR for distance measurement out to about
15 Mpc.

\end{abstract}


\section{Introduction}

With the recent advent of 8-10 meter-class telescopes, it has become
possible for the first time to carry out quantitative spectroscopic
analysis of stellar objects in distant galaxies, opening a new
dimension in the study of the stellar content of such galaxies. These
studies will not only allow us to investigate the stellar properties,
but will also greatly contribute to an understanding of the host
galaxies themselves in terms of star formation properties, and their
chemical and dynamical evolution.  In order to explore this new
opportunity, a few years ago we started a program to carry out
spectroscopy of blue supergiant stars in a number of nearby
galaxies. Blue supergiants are especially well-suited for
spectroscopic analysis in the visual part of the spectrum, because at
visual wavelengths they belong to the brightest objects in a galaxy,
attaining absolute visual magnitudes as bright as $M_V=-10$, thus
pushing the limit for quantitative spectroscopic analysis with
8m-class telescopes out to distance moduli $m-M\simeq 30$. Indeed, we
have recently performed the first quantitative spectroscopic analysis
of a blue supergiant in the galaxy NGC 3621, at a distance of 6.7 Mpc
(Bresolin et al. 2001), more than 100 times the distance to the LMC,
and the most distant galaxy in which quantitative stellar spectroscopy
has yet been carried out. Other galaxies in which blue supergiants
have been studied so far include LMC and SMC (Kudritzki et al.  1989,
Lennon et al. 1991, Puls et al. 1996, Venn 1999, de Koter et al. 1998,
Dufton et al. 2000), the Inter-Cloud population (Rolleston et
al. 1999), M31 (McCarthy et al. 1997, Venn et al. 2000, Smartt et
al. 2001), M33 (McCarthy et al. 1995, Monteverde et al. 1997, 2000),
and NGC 6822 (Muschielok et al. 1999, Venn et al. 2001), all of them
belonging to the Local Group. Going beyond the Local Group, the main
scientific driver behind this work is the possibility to derive
relatively accurate chemical abundances for these stars together with
more accurate estimates of extinction and reddening, even from low
resolution ($R\simeq1000$) optical spectra (Kudritzki 1998), and to
use their wind properties which can be determined from the optical
spectra to get an independent estimate of the host galaxies distances
from the wind momentum-luminosity relationship (WLR, Kudritzki et
al. 1999). Indeed, in the latter paper it was shown that there is
evidence that the WLR for blue supergiants, once properly calibrated
and tested for systematic effects, might yield a standard candle
similar in accuracy to Cepheid variables, and reaching out to the
distance of the Fornax and Virgo clusters.

Before we can fully use this new instrument of distance measurement,
we must thoroughly test its dependence on a number of parameters, most
importantly the stellar spectral type and metallicity. An ideal place
to carry out such a test, and significantly enhance the number of
calibrating objects for the WLR, is the Sculptor Group galaxy
NGC~300. At a distance of $\sim2.0$ Mpc, as derived from Cepheid
variables (Freedman et al.  2001), NGC~300 is close enough to allow
quantitative spectroscopy of its blue supergiant population with
multiobject spectroscopy at the VLT. Furthermore, NGC~300 shows clear
signs of recent massive star forming activity, so a considerable
number of blue supergiants can be expected in this galaxy. Indeed, a
recent wide-field photometric survey of the galaxy carried out by some
of us (Pietrzy{\'n}ski \etal\/ 2001) has identified more than 100 OB
associations. This same survey is currently discovering a large number
of new Cepheids, and the blue supergiant abundances which we will
derive will allow us to determine the abundance gradient in the disk
of NGC~300, from which we hope to obtain the first accurate empirical
determination of the effect of metallicity on the Cepheid
Period-Luminosity (PL) relation, currently only poorly constrained by
observations. The effects of reddening will also be investigated, by
comparing observed and synthetic colors of individual blue
supergiants.

The purpose of this paper is to present our blue supergiant
spectroscopy in NGC~300. In Sec.~2 we describe the target
selection, and the spectroscopic observations and reductions. In
Sec.~3 we present the spectral classification of our targets. In
Sec.~4 we discuss some first results regarding metallicities and
wind momenta, and our conclusions will be presented in Sec.~5.

\section{Blue supergiant selection and spectroscopy}

A large set of multi-epoch, broad-band images have been obtained with
the Wide Field Imager (WFI) at the ESO/MPI 2.2m telescope on La Silla,
as part of a long-term project aiming at the discovery and monitoring
of Cepheids in NGC 300. As mentioned in the Introduction, these data
have already been used by Pietrzy{\'n}ski \etal\/ (2001) to identify
OB associations in NGC~300. Improved {\em BVI} stellar photometry has been
measured for the current work with DAOPHOT/ALLSTAR on a subset (about
20 nights) of the entire dataset, leading to a zero-point accuracy of
$\sim0.03$ mag. For further details on the stellar photometry the reader is
referred to Pietrzy{\'n}ski \etal\/ (2001).

For a preliminary catalog of blue supergiant candidates we selected
all stars brighter than $V=20$, corresponding to an absolute magnitude
$M_V=-6.5$ (luminosity class Iab or brighter for B- and A-type stars)
for an adopted distance modulus $m-M=26.53$ (Freedman et al. 2001),
and having color index in the range $-0.3 < B-V < 0.3$. At this high
galactic latitude ($b=-79^\circ$) the foreground reddening is low,
around $E(B-V)=0.02$ \citep{burstein84}. This, combined with low
internal reddening, makes the observed $B-V$ closely corresponding to
the intrinsic stellar color, and our criterion is therefore optimal
for isolating late B- and early A-type supergiants.  The final list of
candidates for the spectroscopic follow-up, containing 167 objects,
was set up by carefully examining the original WFI frames, rejecting
objects on the basis of broad profiles and presence of nearby
companions. An H$\alpha$ image of the galaxy was also inspected in
order to avoid overlap with emission line nebulae.

Spectroscopy of a subset of our candidate list was obtained with
Antu and FORS1 at the Very Large Telescope (Paranal) on September
25 and 26, 2000.  Two FORS1 fields were observed each night in
multi-object spectroscopy mode, allowing simultaneous spectroscopy
of 19 objects, for a total of four different pointings, chosen to
allow a good coverage of the radial extent of the galaxy.  Sky and
seeing conditions were excellent on both nights, with typical 0.7
arcsec seeing, but with long spells of 0.4--0.5 arcsec seeing.
Five exposures, each lasting 45 min, were secured at every
pointing with a 600 gr/mm grating, which provides approximately a
5~\AA\/ spectral resolution. The spectral coverage with this
setup is about 1,000 \AA\/ wide, centered around 4,500 \AA\/
(dependent on an object's position in the focal field along the
dispersion axis), including in most cases the range from the H
and K calcium lines up to the Balmer H$\beta$ line.

Due to the positioning limitations of the FORS slitlets and the uneven
distribution of blue supergiants in NGC 300, a few additional objects,
not included in our candidate list, were also observed. Among these
were H~II regions, blue stars somewhat fainter than our original
magnitude limit, and a handful of late-type stars. Central coordinates
of the fields observed are reported in Table~1, while
Fig.~\ref{galaxy} shows the location of these fields on a wide-field
image of NGC 300. The individual fields, together with the
identification of the spectroscopic targets, are shown in
Fig.~\ref{fieldA} through \ref{fieldD}, reproducing $V$-band, 5-min
FORS1 exposures.  Table~2 summarizes the positions, {\em BVI} magnitudes
and additional information for all the objects for which a spectrum
was obtained. In this Table and for the rest of this paper, individual
stars will be identified with the letter corresponding to the galaxy
field (A through D) and the progressive FORS slitlet number (1 through
19).

The generic image processing tasks within IRAF\footnotemark
\footnotetext[4]{IRAF is distributed by the National Optical Astronomy
Observatories, which are operated by AURA, Inc., under contract with
the National Science Foundation.} were used for bias and flat field
corrections. After adding all the images of a given field, each
individual slitlet spectrum was treated as a long-slit spectrum, and
independently wavelength calibrated and extracted.  Finally, the 1-D
spectra were normalized with a low-order polynomial.  Our targets are
mostly located in uncrowded regions, so that sky subtraction did not
pose particular problems.  The only difficult cases were represented
by stars in the proximity of or within emission nebulae. Despite our
efforts to avoid such occurrences by using the H$\alpha$ image of
NGC~300, the spectra of several stars were found to be contaminated by
nebular emission. At this spectral resolution a complete and
satisfactory subtraction of this contamination is not possible, and we
marked the affected objects in Table~2. The average S/N for most of
the spectra is close to 50, but for the brightest stars it goes up to
$\simeq100$. Only very few spectra are underexposed (S/N $<$ 25).

\section{Spectral classification}

The spectral classification of our targets was carried out by a
visual comparison of the observed spectra with template spectra of
B- and A-type Galactic supergiants, and B-type supergiants in the
SMC, degraded to FORS resolution. While the SMC data have been
taken from the literature (Lennon 1997), the Galactic data are
part of an ongoing project, which aims at obtaining
high-resolution spectra of nearby blue supergiants for accurate
stellar atmospheric analysis. The overall appearance of the
available spectra compared with the template spectra was used for
the spectral classification, with special attention to widely-used
diagnostics of blue supergiants.

As pointed out by Lennon (1997), ambiguities in the spectral
classification of extragalactic B-supergiants can arise if the
classification scheme does not account for possible significant
deviations from galactic metallicity. We have therefore applied his
classification criteria, which are more independent of metallicity,
and have used both sets of templates, Galactic and SMC, for B-type
spectra. However, in a few cases ambiguities remained, which we will
be able to eliminate only after a detailed quantitative analysis of
the spectra has been carried out. Those cases are marked in Table~2
and the comparison shows a relatively small effect (a shift of one or
two subclasses) due to the different metallicity of the template
spectra.

The spectral types thus determined are presented in column 7 of
Table~2. In several cases we provide a range of spectral classes,
reflecting the uncertainty due to the S/N of the spectra and the
possible abundance effects. Additional comments pertaining to the
appearance of the spectra or the spectral classification are
given in column 8. Objects which show a possibly composite
spectrum were not assigned a spectral classification. The nebular
contamination has been identified as such when the 2-D spectra
indicated the presence of emission lines extending over and beyond
the stellar position.

\begin{figure}
\plotone{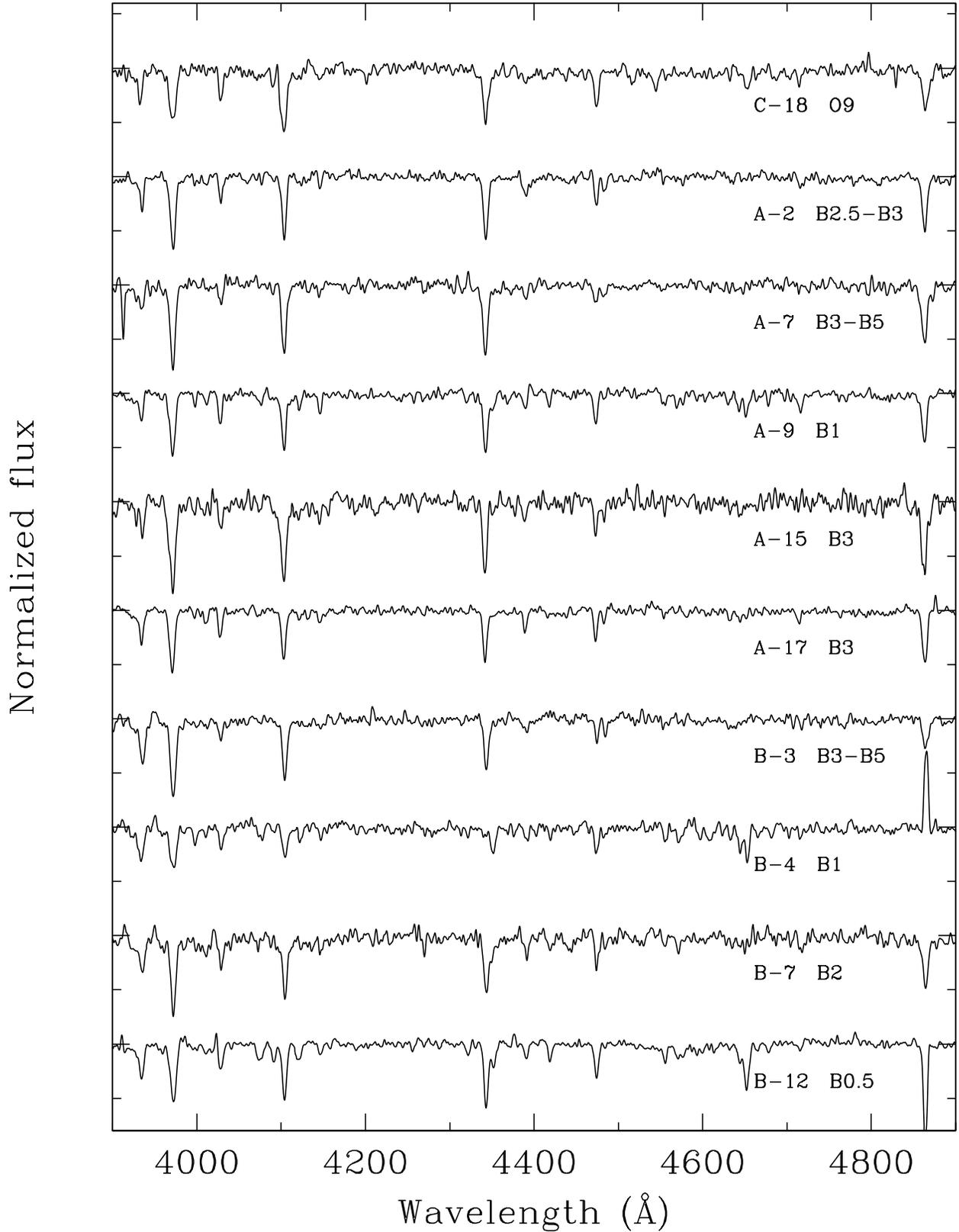} \caption{Spectral catalog of early-B
supergiant stars in NGC 300. The first object is an O9 star. In this
and all following figures the spectra are not displayed in the NGC~300
rest frame.}\label{catalog1}
\end{figure}

\begin{figure}
\plotone{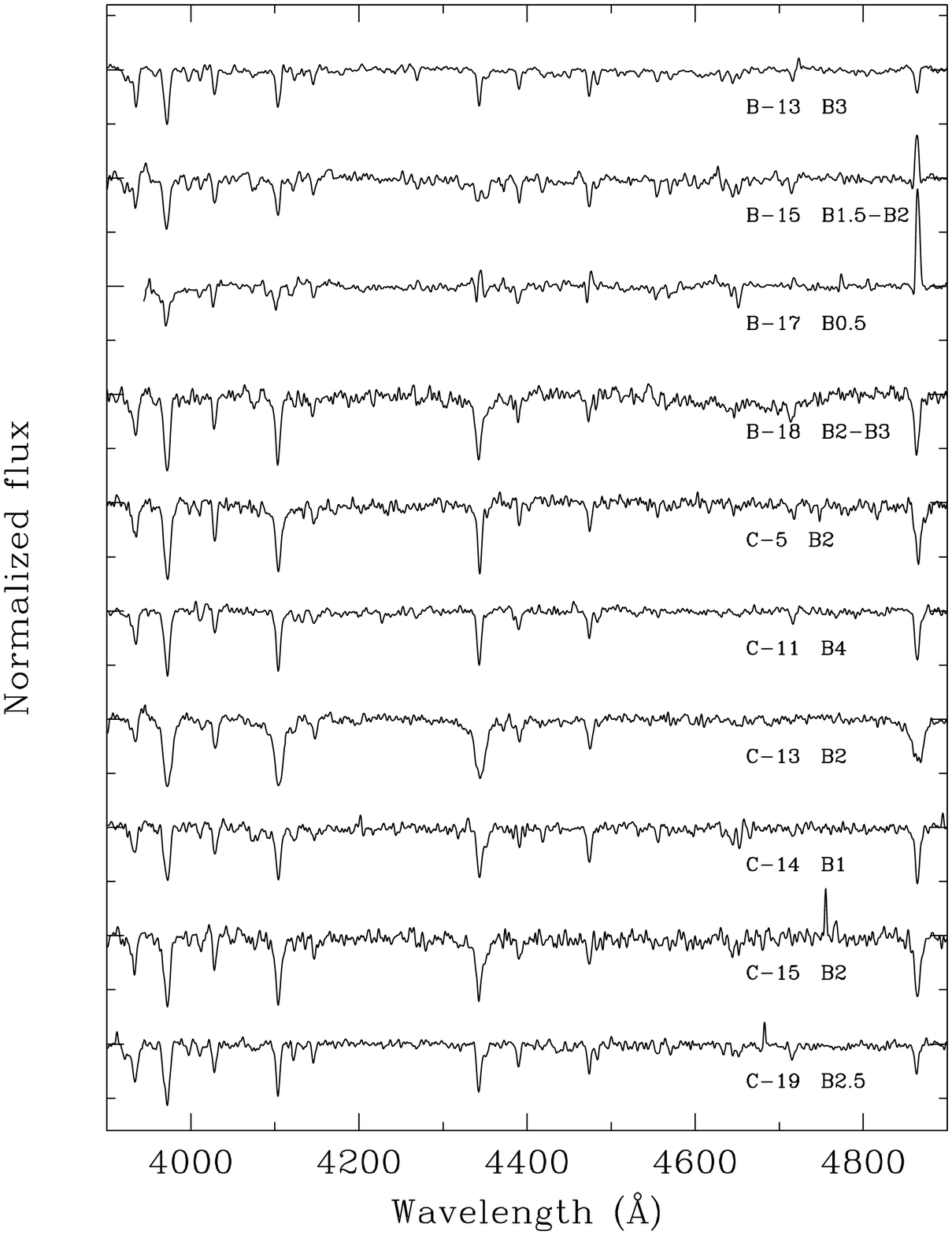} \caption{Spectral catalog of early-B
supergiant stars in NGC 300.}\label{catalog2}
\end{figure}

\begin{figure}
\plotone{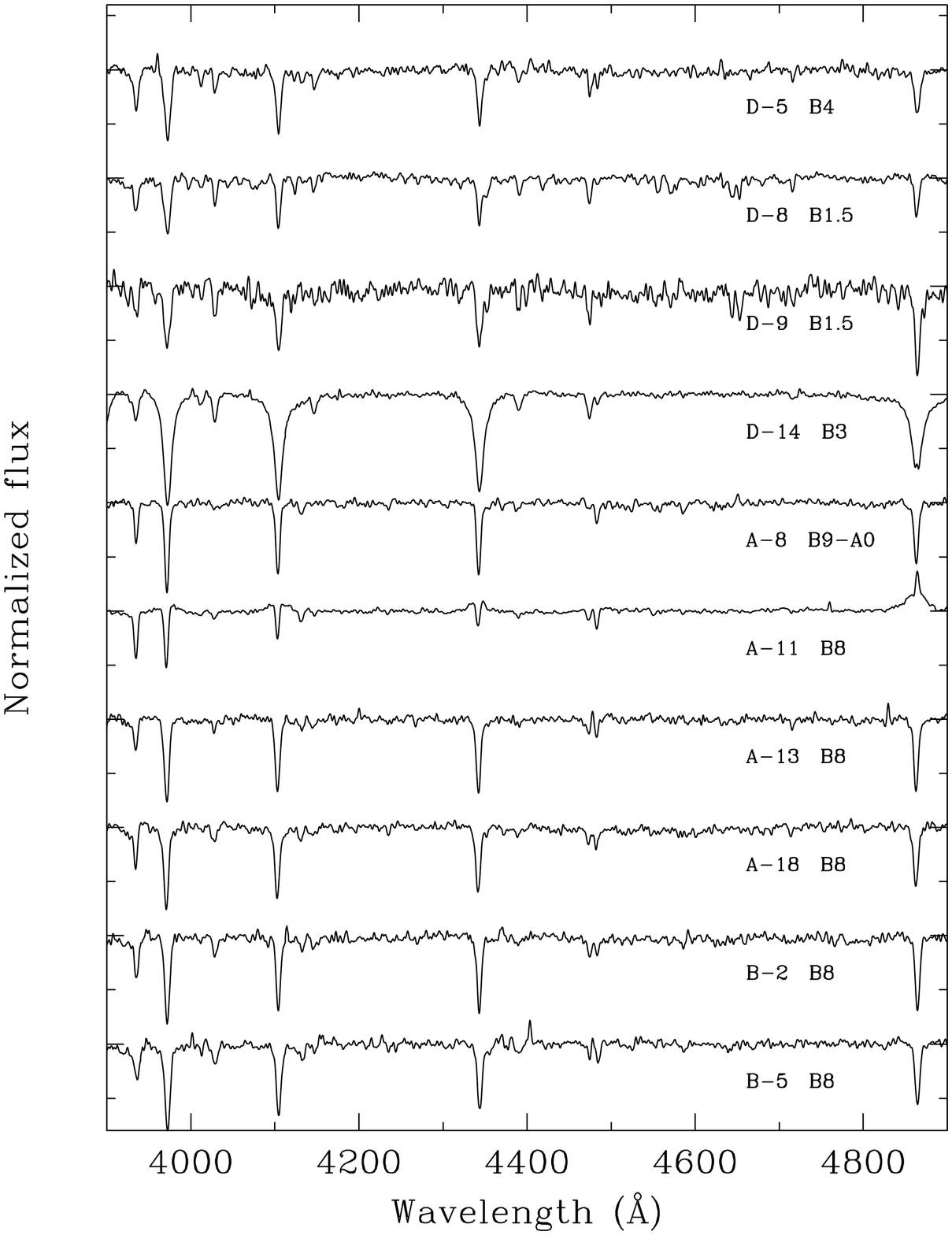} \caption{Spectral catalog of early- and
late-B supergiant stars in NGC 300.}\label{catalog3}
\end{figure}

\begin{figure}
\plotone{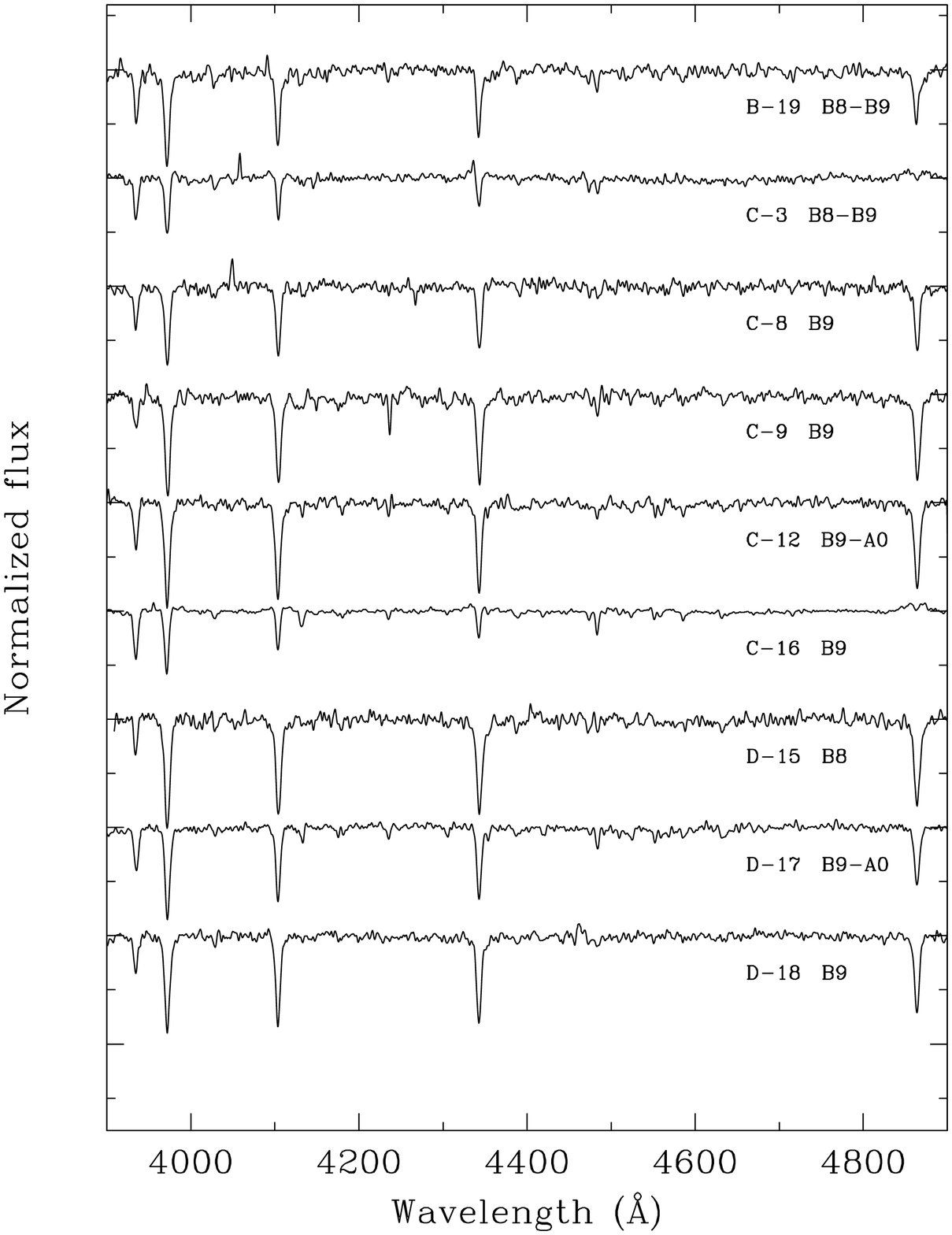} \caption{Spectral catalog of late-B
supergiant stars in NGC 300.}\label{catalog4}
\end{figure}

\begin{figure}
\plotone{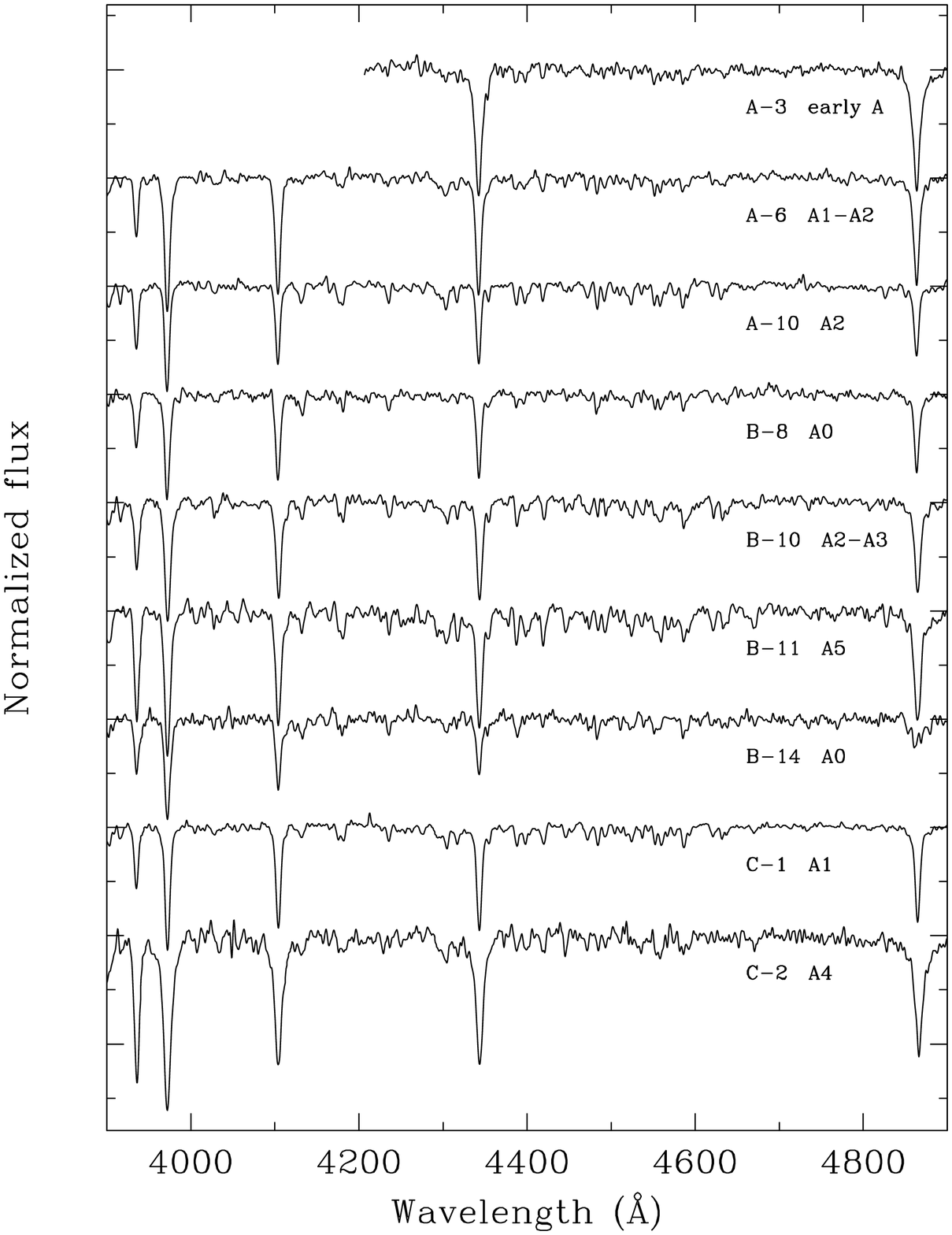} \caption{Spectral catalog of early-A
supergiant stars in NGC 300.}\label{catalog5}
\end{figure}

\begin{figure}
\plotone{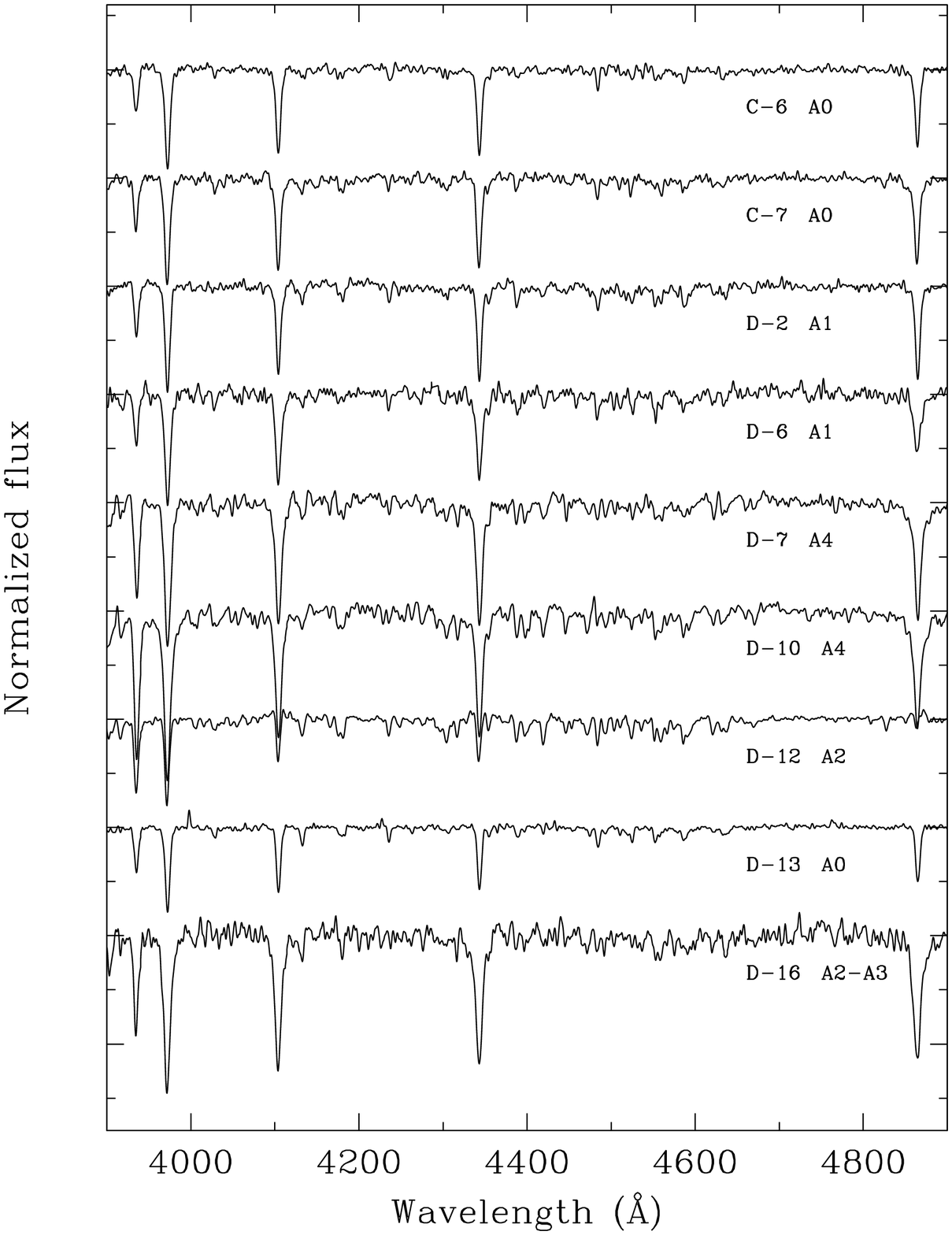} \caption{Spectral catalog of early-A
supergiant stars in NGC 300.}\label{catalog6}
\end{figure}

\begin{figure}
\plotone{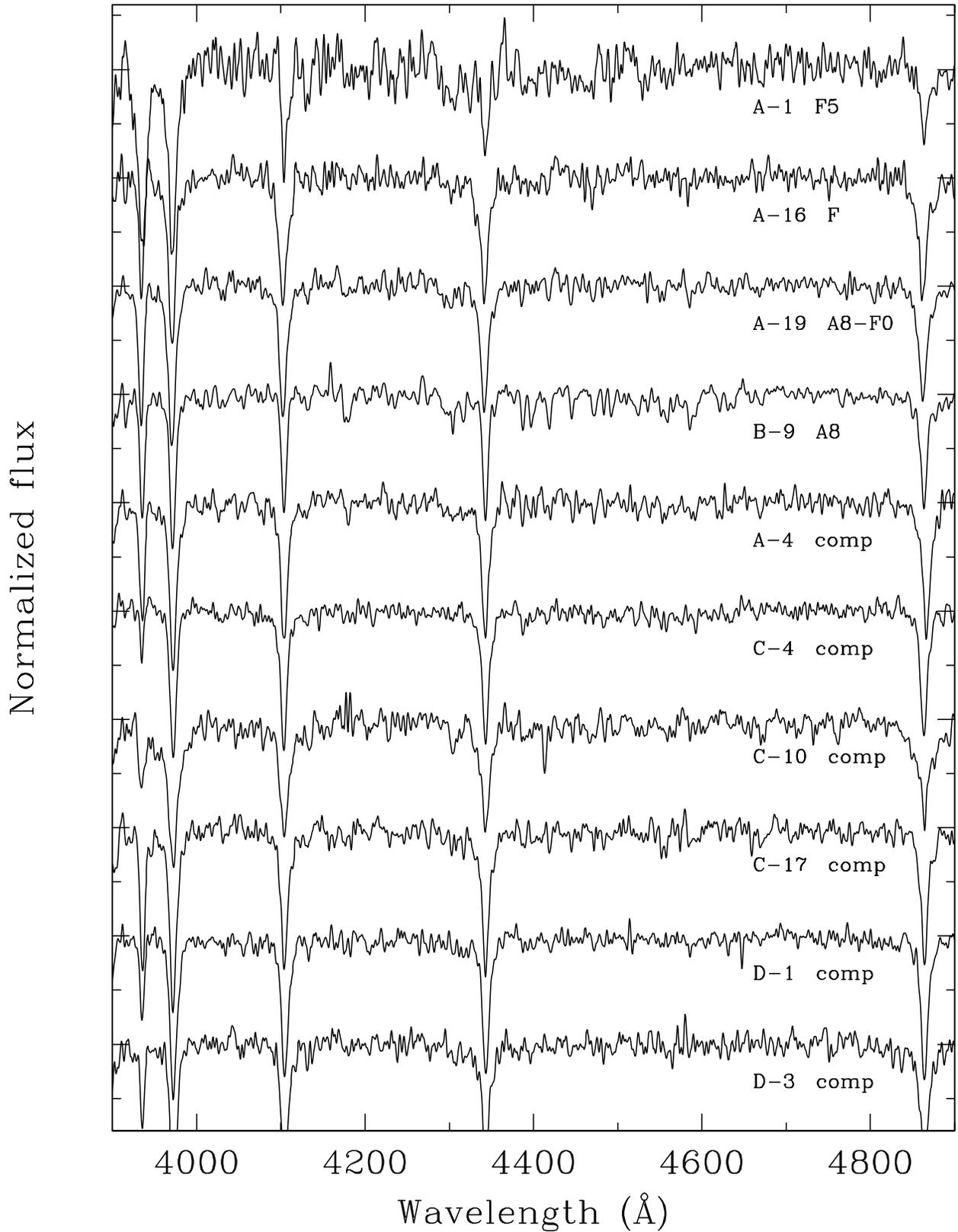} \caption{Spectral catalog of late-A and F supergiant
stars in NGC 300, and several possibly composite spectra.}\label{catalog7}
\end{figure}

We present in Figs.~\ref{catalog1}-\ref{catalog7} the spectra of the
NGC~300 supergiants, grouped by spectral type as follows: late-O and
early-B (B0 through B5), late-B, early-A (A0 through A5), late-A and F
stars.  We exclude from this spectral atlas the H~II region spectra,
and some additional objects, which include a foreground white dwarf
and a WN11 star, which will be discussed in a future paper.

\begin{figure}
\plotone{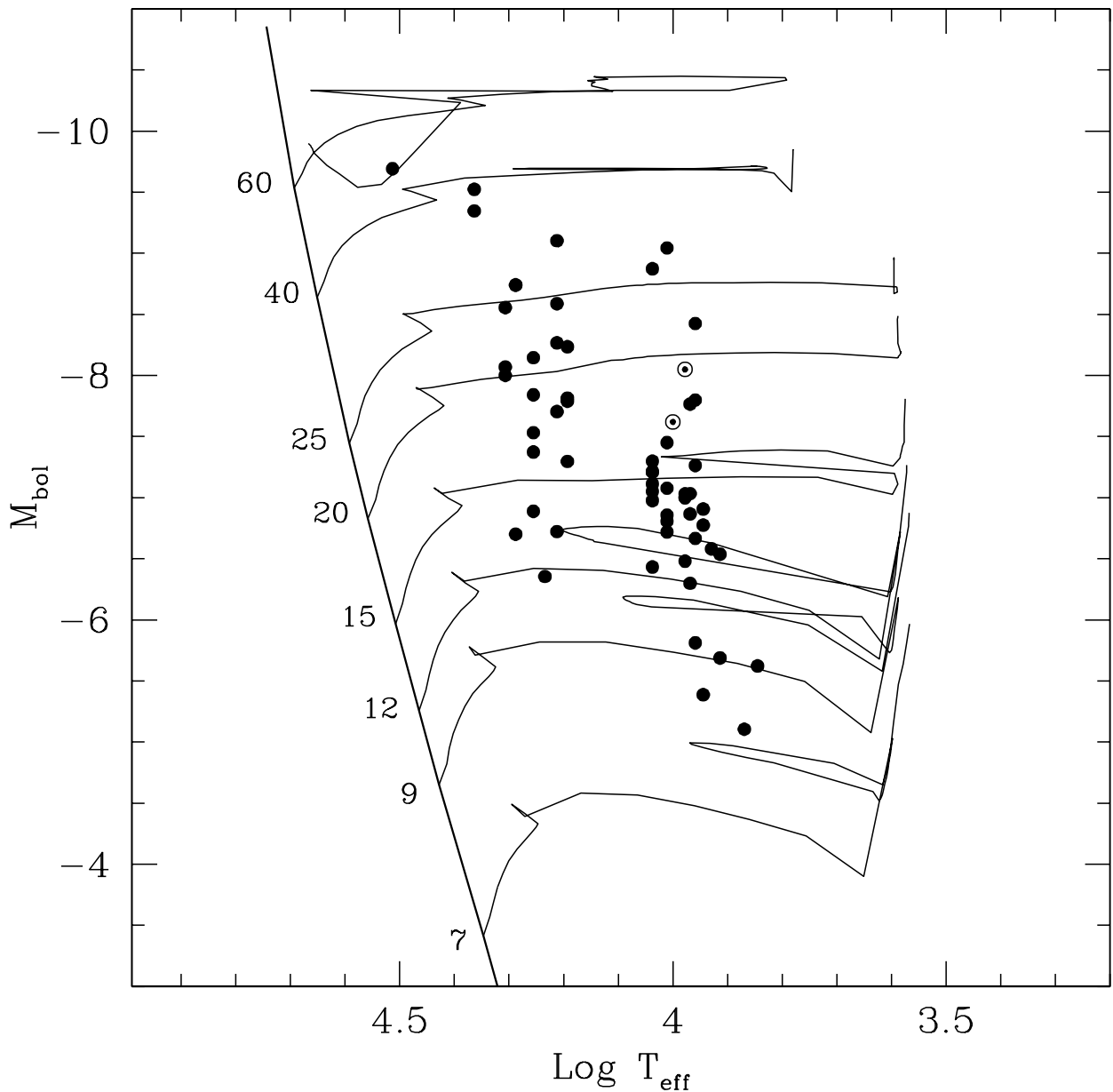} \caption{$M_{bol}$ vs. \teff\/ diagram of the blue
supergiants observed in NGC~300. The model tracks at metallicity
Z=0.008 of Schaerer et al. (1993) for different initial stellar masses
are shown, as indicated by the numbers to the left of the Main
Sequence. The gap around $\log$\,\teff\,=4.1 is due to the
unavailability of spectral templates between types B5 and B8 in our
spectral classification program. The two stars described in Sec. 4.1
are shown with different symbols.}\label{hr}
\end{figure}

The position of the supergiant sample in the HRD is shown in
Fig.~\ref{hr}.  The stellar luminosity was derived after correcting
for reddening (assuming $A_{V} = 3.1\,E(B-V)$). The latter was
obtained from the observed $B-V$ and the expected value of this color
index for a given spectral type, as given by Fitzgerald
(1970). Bolometric corrections and effective temperatures as a
function of spectral type were taken from Humphreys \& McElroy (1984).
Theoretical stellar tracks at metallicity Z=0.008 from Schaerer et
al. (1993) are also shown for comparison (the new tracks which include
stellar rotation by the Geneva group and used in Sect.~4.1 to estimate
the stellar parameters of two bright supergiants are not yet available
for this metallicity). The apparent gap centered around
$\log$\,\teff\,=4.1 is the result of a selection effect, due to the
lack of template spectra between types B5 and B8 in our classification
program.

\section{First results}

\subsection{Stellar parameters and abundances}

While at the resolution currently attainable in multiobject
spectroscopy of targets as faint as our blue supergiants a detailed
chemical abundance analysis is difficult (though not impossible), we
can in any case estimate, within roughly 0.2 dex, the abundances of
several elements by a comparison of the observed spectra to synthetic
ones generated by models of blue supergiant atmospheres, calculated
for a variety of metal abundances. We leave a full abundance analysis
of our blue supergiant sample to a forthcoming paper, but we want to
show here some first results illustrating the power of our technique
to estimate the metal abundances of these stars.

The photospheric analyses are performed on the basis of hydrostatic
LTE line-blanketed model atmospheres (Kurucz 1991) and subsequent
non-LTE/LTE spectrum synthesis. Effective temperatures \teff\/
are estimated from the spectral classification, and surface gravities
$\log g$ are determined from the Balmer line strengths; the
microturbulence $\xi$ is assumed to be the same as in the Galactic
comparisons. The helium content $y$ (by number) and the stellar
metallicity $[$M/H$]$ are deduced from the comparison of the observed
and the synthetic spectra for varying elemental abundances. At
present, the elements with strong lines in A-type stars, e.g. Mg\,{\sc
ii}, Ti\,{\sc ii} and Fe\,{\sc ii}, are treated in non-LTE, while
Si\,{\sc ii} and Cr\,{\sc ii} are treated in LTE. In total, several
ten-thousand lines from these and some 20 other chemical species --
comprising almost all spectral features -- are included in the
spectrum synthesis (with the CNO and S lines also in non-LTE). The
results from the analysis of two of the NGC~300 supergiants, A-8 and
D-13, are summarized in Table~3.  Radial velocities $v_{\rm rad}$ can
be determined from cross-correlation of the observed spectra with the
synthetic ones.

\begin{figure}
\plotone{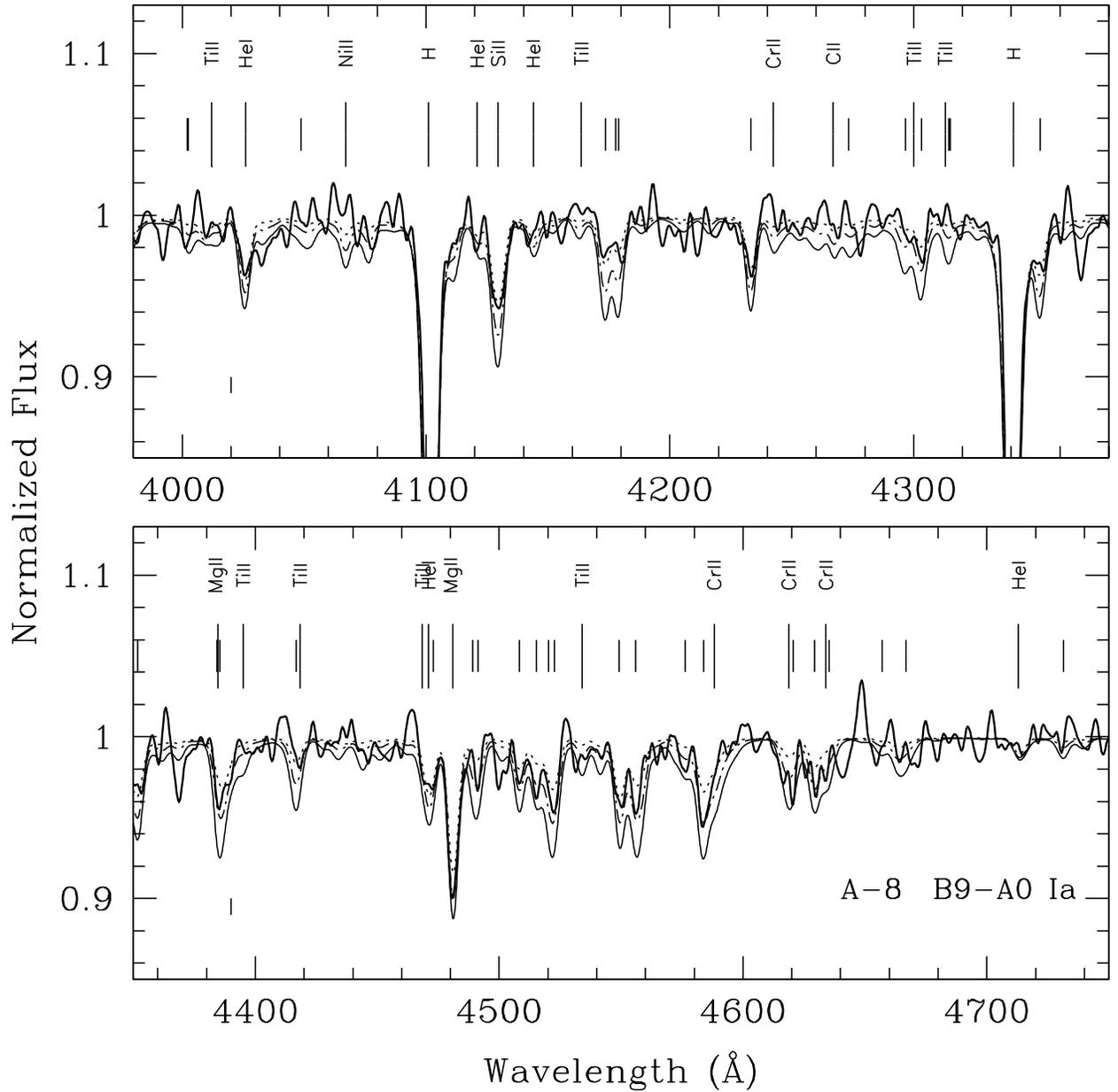} \caption{Model fits, corresponding to metal
abundances of 0.2, 0.5 and 1.0 solar (dotted, dashed and solid lines,
respectively), to the spectrum of the B9-A0 supergiant A-8 in NGC~300
(thick solid line). Short vertical marks in the element identification
section indicate Fe\,{\sc ii} lines.  The metal abundance of this star
is $\sim0.2$ solar. The estimated 1\% noise level in the continuum is
shown by the bar at the lower left of each plot.}\label{a8}
\end{figure}

\begin{figure}
\plotone{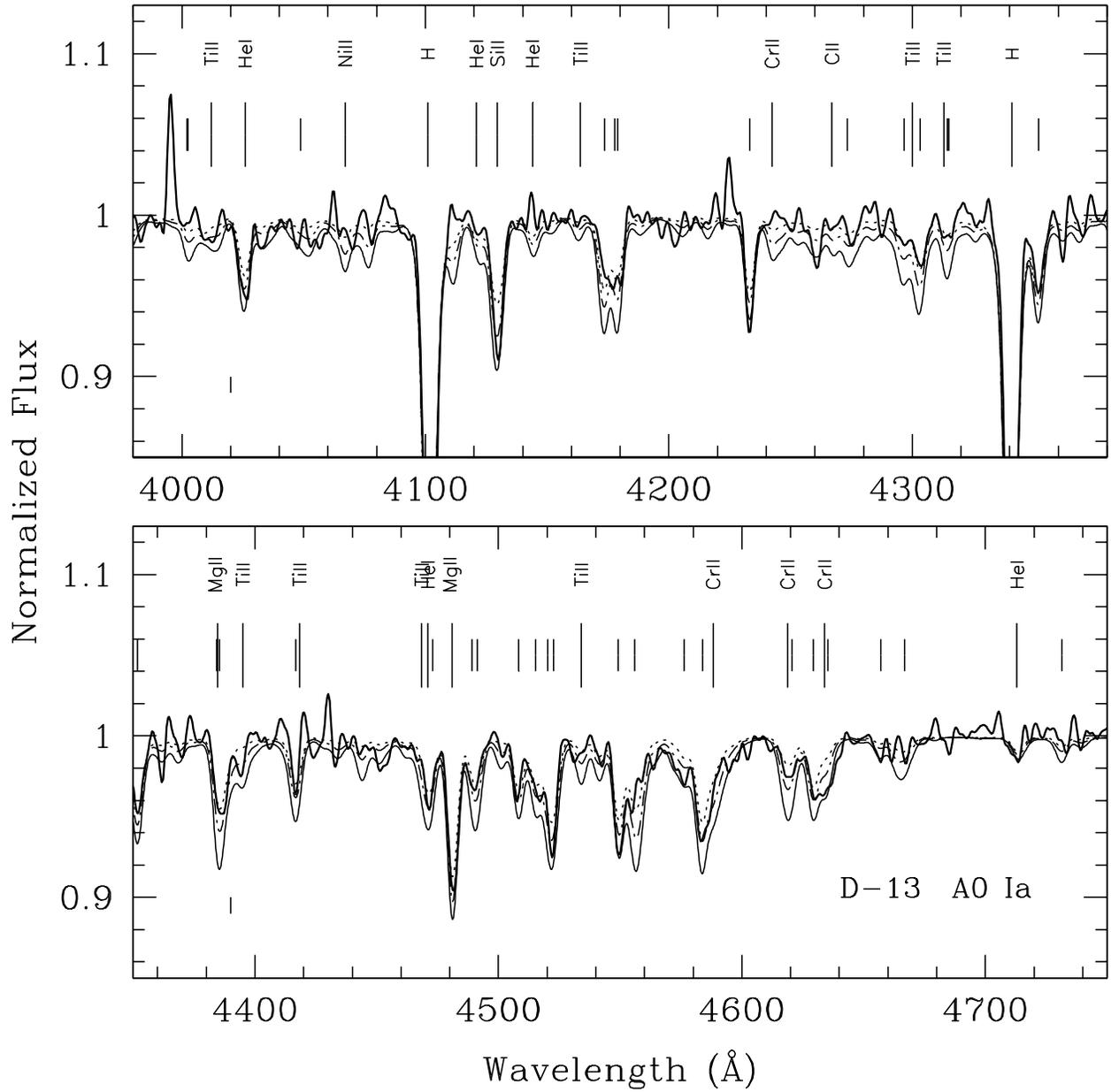} \caption{As Fig. 13, for the A0 supergiant D-13. The
metal abundance of this star is higher than that of A-8, $\sim0.5$
solar.}\label{d13}
\end{figure}

In Fig.~\ref{a8} we show the observed spectrum of the A0 Ia star A-8,
together with model fits which were calculated for three different
metal abundances: 0.2, 0.5 and 1.0 solar. A comparison of the
predicted and observed line intensities, particularly for lines of
elements Fe and Cr, suggests a low abundance for this star, in the
order of 0.2 solar. This is consistent with the star's position in the
outskirts of the galaxy, where a low metal abundance would be
expected. In Fig.~\ref{d13} we show the same model fits to the
spectrum of another A0 supergiant, star D-13. This object has a
clearly higher metal abundance, in the order of 0.5 solar. Again this
is consistent with the expected higher metallicity for this star,
which is considerably closer to the center of NGC~300.  In both
figures the estimated noise level in the continuum (1\%) is shown by
vertical bars at the lower left. An additional 1\% uncertainty is
estimated for the placement of the continuum. The latter was measured
from low-order polynomial fits to relatively line-free regions of the
observed spectra, where the theoretical normalized fluxes predicted
for different metallicities reach unity simultaneously (e.g. around
4160\,\AA, 4200\,\AA, 4610\,\AA, 4690\,\AA). This is contrast with the
situation in the UV part of the spectrum, where the `true' continuum
is never observed (Haser et al. 1998).

Applying the technique described above to the complete sample of
supergiants, we expect to delineate the radial metallicity gradient in
the disk of NGC~300 rather accurately. It will be of considerable
interest to compare the stellar abundance gradient to the one derived
from H~II region abundance studies, which in extragalactic work has
been assumed to reflect the stellar abundances, e.g. in the work of
the HST Key Project team on M101 to measure the metallicity effect on
the Cepheid PL relation (Kennicutt et al. 1998).

Having determined the atmospheric parameters, the physical properties
of the stars are derived, cf. Table~3. The reddening $E(B-V)$ is found
from the comparison of the photometry with the synthetic colors from
the model fluxes.  Absolute visual magnitudes $M_{V}$ are obtained
after correcting for extinction, assuming $A_{V} =
3.1\,E(B-V)$. Applying a bolometric correction $B.C.$ leads to the
stellar bolometric magnitude $M_{bol}$.  From this and the atmospheric
parameters \teff\/ and $g$, the stellar luminosity $L$, the radius $R$
and the spectroscopic mass $M^{\rm spec}$ are determined. Zero-age
main sequence masses $M^{\rm ZAMS}$ are derived from comparison with
stellar evolution tracks accounting for mass-loss and rotation (Meynet
\& Maeder 2000; Maeder \& Meynet 2001).  The supergiant A-8 is
situated in a region of the HRD where partial blue loops may be found
in stellar evolution calculations, depending on the detailed physics
accounted for (see discussion in Maeder \& Meynet 2001). An enhanced
He abundance supports such an interpretation for the evolutionary
status of this object (cf. the 12 $M_\odot$ track of Maeder \& Meynet
2001).  The more massive supergiant D-13 has either developed
directly from the main sequence (in the case of an initially fast
rotator) or has reached the post red supergiant phase (as an initially
slow rotator), accounting for its marked helium enhancement and its
spectroscopic mass (cf. the 20 $M_\odot$ track of Meynet \& Maeder
2000). Note that for this metallicity sophisticated stellar evolution
tracks accounting for rotation are not available yet, but those for
solar metallicity are expected to be sufficiently similar.
Unfortunately, information on the N/C ratio, the most sensitive
indicator on the evolutionary status, cannot be derived from the
available spectra.

To conclude, the wealth of data obtained on just two sample
supergiants clearly demonstrates the versatility of quantitative
spectroscopy for stellar and -- through the observation of a larger
ensemble of objects -- galactic studies.

\subsection{Wind Momentum-Luminosity Relationship (WLR)}

The existence of a relationship between the stellar wind momentum
and the luminosity of hot massive stars is a sound prediction of
the theory of radiatively driven winds (Kudritzki 1998, Kudritzki
\& Puls 2000). The relationship has the form

$$\dot{M}v_\infty \propto R^{-0.5} L^{1/\alpha}$$

where the product of mass-loss rate ($\dot{M}$) and the wind terminal
velocity ($v_\infty$) gives the mechanical momentum flow carried away
by the stellar wind. $R$ is the stellar radius, $L$ the luminosity,
and $\alpha$ is the exponent of the power-law line strength
distribution function of the metal lines driving the wind. As the
ionizing properties in the stellar atmosphere change with effective
temperature, so do the line strengths of the metal lines most
effective at driving the wind, and as a consequence $\alpha$ is
expected to vary with stellar spectral type, and values from
$\sim0.65$ (O-type) to $\sim0.38$ (A-type) are found in the solar
neighborhood (Kudritzki \& Puls 2000).

\begin{figure}
\plotone{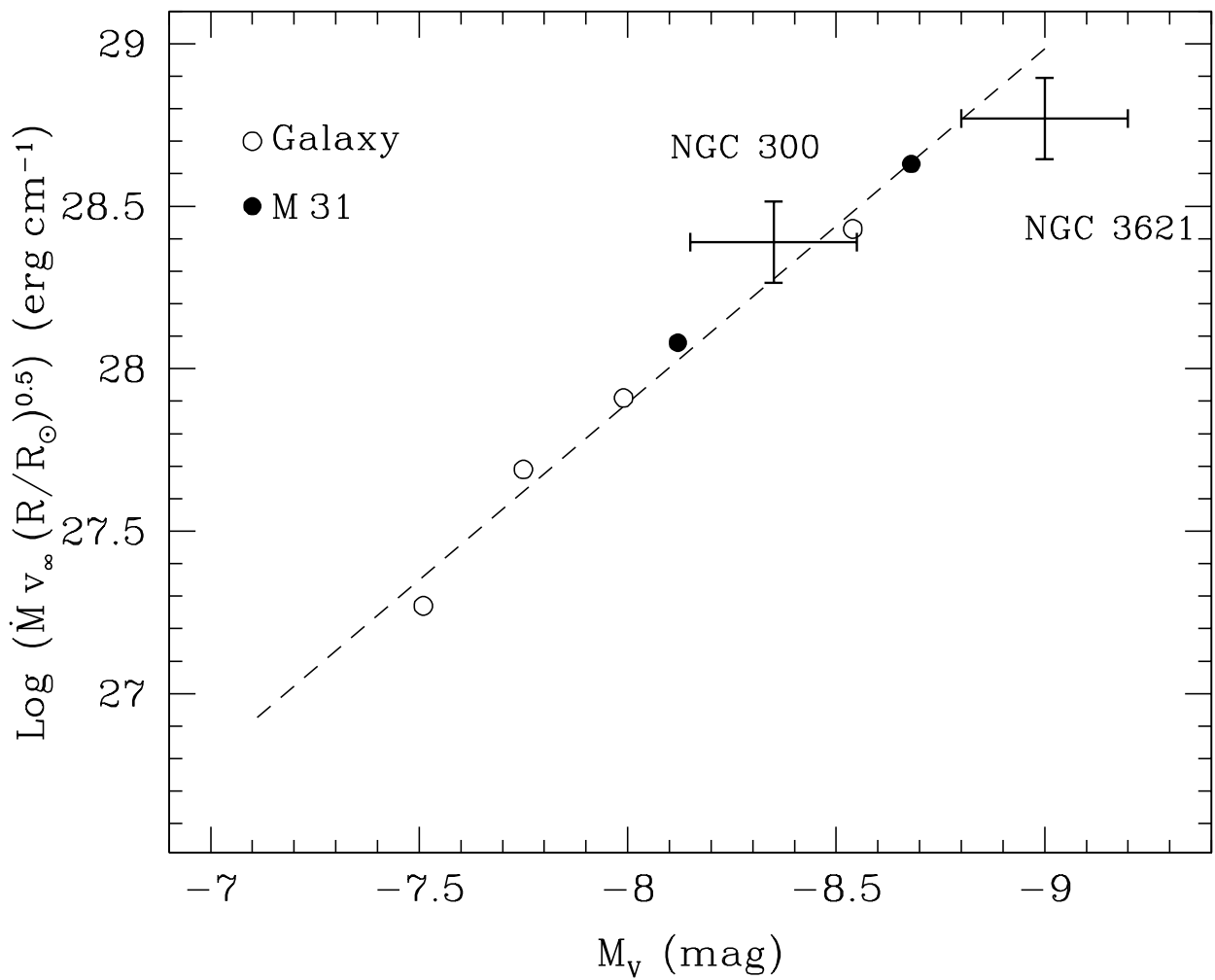} \caption{Wind momentum-luminosity relationship for
A-type supergiants taken from data of Galactic (open circles) and M31
(filled circles) supergiants (data taken from Kudritzki et al. 1999;
improved photometry and distance modulus were used for M31, see
Bresolin et al. 2001).  Also plotted is the A0 supergiant examined by
Bresolin et al. (2001) in NGC 3621, and star D-12 in NGC~300 analyzed
in this paper.}\label{wlr}
\end{figure}

The validity of the WLR has been demonstrated empirically by Puls et
al. (1996) for O-type stars in the Galaxy and the Magellanic Clouds,
as well as for Galactic supergiants of type B and A (Kudritzki et
al. 1999), confirming the expected dependence of the relation on
spectral type, roughly in agreement with the slopes predicted by
theory. Our new, large sample of blue supergiants in NGC~300, all at a
given distance (which will eventually be improved with the Cepheids we
are currently detecting in a parallel program) provides an extremely
valuable dataset to improve the calibration of the WLR and establish
its dependence on spectral type and metallicity, with a much higher
accuracy than hitherto possible.  The most accurate results for the
wind momentum of the blue supergiants can be obtained from a modeling
of the Balmer H$\alpha$ line profile (for a review, see Kudritzki \&
Puls 2000), and we are looking forward to getting red spectra covering
H$\alpha$ for our complete sample of blue supergiants with VLT later
this year. However, for the brightest stars it is possible to derive
the wind momentum also from the H$\beta$ line, albeit with lower
accuracy. We have done this for the A2 supergiant D-12. This star has
an absolute visual magnitude $M_V=-8.35$, and is one of the brightest
supergiants in our sample. From a fit to the H$\beta$ line profile,
the mass-loss rate was obtained ($1.8\,(\pm 0.2)\,\times10^{-6}$
M$_\odot$/yr), which, together with an assumed value for $v_\infty$
(150 km/s, a typical value for A2 supergiants, cf. Lamers et al. 1995
and Kudritzki \& Puls 2000) and radius (210 $R_{\odot}$, derived from
$E(B-V)$ = 0.15, $M_V = -8.35$ and model atmosphere flux), yielded the
position in the diagnostic diagram shown in Fig.~\ref{wlr}. It is seen
that this `preliminary' datapoint from this one supergiant fits
relatively well into the existing WLR for A supergiants, making us
optimistic that we can achieve a significant improvement on the WLR
with the results based on the H$\alpha$ profiles.

\section{Conclusions}

We have presented a spectral catalog of about 70 blue supergiant
candidates in NGC~300, observed at a resolution $R\simeq1000$ and S/N
$\simeq50$ in the 4,000-5,000~\AA\/ spectral range at the VLT. Of the
observed targets, 62 are spectroscopically confirmed as supergiants
with spectral types between late-O and F. Most of these supergiants
are types B and A. In our survey, we also found several different,
interesting objects including a WN11 star and a foreground white
dwarf, which will be analyzed in detail in forthcoming studies. The
spectral classification of the blue supergiants determined in this
paper will be essential for a thorough investigation of the dependence
of the wind momentum-luminosity relationship as a new, purely
spectroscopic and far-reaching distance indicator on spectral type, as
predicted by theory, and seen in preliminary empirical results
(Kudritzki et al. 1999).

A model atmosphere technique was employed to obtain first results for
the metal abundances of two A0 supergiants in our sample.  A
comparison of synthetic spectra, calculated for different metal
abundances, with the observed spectra of these stars, yields metal
abundances different by approx 0.3 dex in the expected sense that the
star closer to the center of NGC~300 is more metal-rich than its
counterpart, which is located at a larger galactocentric distance. In
a forthcoming paper we will use this technique on the whole sample of
blue supergiants in NGC~300 to determine their metal abundances. While
the individual values of these abundances will be of a modest accuracy
(about $\pm0.2$ dex per star), the large number of stars in our sample
and the wide range in galactocentric distance they span, should allow
us to determine the abundance gradient in the disk of NGC~300 with an
accuracy which is unprecedented in the study of spiral galaxies beyond
the Local Group.

We also report on a first determination of the wind momentum for one
of the brightest A-type supergiants in our sample, based on its
mass-loss rate as determined from the Balmer H$\beta$ line profile,
and find that it fits relatively well into the empirical WLR
determined from A supergiants in the Galaxy and M31 by Kudritzki et
al. (1999). We will be able to obtain the wind parameters with higher
accuracy from H$\alpha$ profiles which we expect to have at our
disposal, for all stars in our sample, by the end of 2001.  We will
then be able to perform a thorough empirical check of the usefulness
of the WLR for distance determinations, including a calibration of its
dependence on spectral type and metallicity, and of its intrinsic
dispersion, for a given spectral type and metallicity.

\acknowledgments

FB acknowledges DLR grant 50 OR 9909 for support while working in
Munich.  WG gratefully acknowledges financial support received from
Fondecyt grants 1000330 and 8000002 for this project. Part of this
work was done while he was a scientific visitor at ESO Garching. WG is
grateful for the support received from ESO. We also thank the referee,
A. de Koter, for positive and constructing comments.

\clearpage

\begin{figure}
\caption{The four FORS1 fields observed in NGC 300 are marked on a montage of eight B-band ESO/MPI\,
2.2m~+~WFI frames. North is at the top, east to the left. Field size
is approximately $34\arcmin \times 33\arcmin$.}\label{galaxy}
\end{figure}

\begin{figure}
\caption{Field A from a 5-min, V-band FORS1 exposure. The field of
view is approximately $6.8\arcmin \times 6.8\arcmin$. The multi-object
spectroscopy targets are marked by the circles. Their identification
numbers correspond to those in Table~2.}\label{fieldA}
\end{figure}

\begin{figure}
\caption{Field B. The multi-object spectroscopy targets are marked by
the circles. A bright foreground star has been masked out using one
arm of the FORS MOS unit.}\label{fieldB}
\end{figure}

\begin{figure}
\caption{Field C. The multi-object spectroscopy targets are marked by
the circles.}\label{fieldC}
\end{figure}

\begin{figure}
\caption{Field D. The multi-object spectroscopy targets are marked by
the circles. A bright foreground star has been masked out using two
arms of the FORS MOS unit.}\label{fieldD}
\end{figure}

\clearpage
\begin{deluxetable}{ccccc}
\tabletypesize{\scriptsize}
\tablecolumns{5}
\tablewidth{0pt}
\tablecaption{NGC 300 - Observed fields}

\tablehead{
\colhead{Field}     &
\colhead{RA (2000.0)}            &
\colhead{DEC (2000.0)}           &
\colhead{Position Angle (deg)}    &
\colhead{Obs. date}}
\startdata
A & 00 55 36.4 & $-$37 41 36.5 & \phantom{-13}0 & Sep. 25, 2000 \\
B & 00 54 52.1 & $-$37 40 26.9 & \phantom{-13}0 & Sep. 25, 2000 \\
C & 00 54 32.4 & $-$37 36 16.7 & \phantom{-1}20 & Sep. 26, 2000 \\
D & 00 54 30.6 & $-$37 42 21.7 & $-$13 & Sep. 26, 2000 \\
\enddata
\end{deluxetable}

\clearpage
\begin{deluxetable}{rrrrrrll}
\tabletypesize{\scriptsize}
\tablecolumns{8}
\tablewidth{17.5cm}
\tablecaption{NGC 300 - Spectroscopic targets}

\tablehead{
\colhead{Slit no.}     &
\colhead{RA}            &
\colhead{DEC}           &
\colhead{$V$}             &
\colhead{$B-V$}           &
\colhead{$V-I$}     &
\colhead{Spec. type}    &
\colhead{Comments}}
\startdata

FIELD A & & & & & & & \\
1   & 0 55 37.069  & $-$37 38 19.44 &   21.50  &  0.49   &  0.60 & F5  & noisy\\
2   & 0 55 36.674  & $-$37 38 31.03 &   20.04  & $-$0.11  & $-$0.05 & B2.5--B3  & Balmer lines stronger than standards\\
3   & 0 55 47.220  & $-$37 38 50.51 &   20.10  &  0.07  &   0.21 & early A  & spectrum incomplete, Balmer lines\\
& & & & & & & strong, composite?\\
4   & 0 55 22.188  & $-$37 39 17.08 &   20.86  & 0.12    & 0.31 &   & noisy, Balmer lines strong, composite?\\
5   & 0 55 32.104  & $-$37 39 39.30 &   20.71  &  2.18   & \nodata &  & very noisy \\
6   & 0 55 37.828  & $-$37 40 07.01 &   19.74  &  0.03   &  0.12 & A1--A2 & Si\,{\sc ii} 4128/32 missing, Mg\,{\sc ii} 4481 weak,\\
& & & & & & & metal weak? Balmer lines strong \\
7   & 0 55 37.865  & $-$37 40 23.20 &   20.43  & $-$0.11   & 0.10 & B3--B5  & for Gal. abundance; B5--B8 if metal-poor\\
8   & 0 55 41.619  & $-$37 40 58.61 &   19.44  & $-$0.01 & \nodata & B9--A0  & \\
9   & 0 55 39.262  & $-$37 41 18.23 &   20.23  & $-$0.17 & \nodata & B1  & \\
10  & 0 55 31.961  & $-$37 41 38.33 &   18.98  &  0.07 & \nodata & A2  & \\
11  & 0 55 38.317  & $-$37 41 56.81 &   18.41  & 0.07 & \nodata & B8  & emission wings in Balmer lines, H$\beta$\\
& & & & & & & in emission, very strong wind?\\
12  & 0 55 32.497  & $-$37 42 26.62 &   19.49  &  0.14  &  $-$0.06 &   & foreground white dwarf\\
13  & 0 55 33.276  & $-$37 42 46.78 &   19.83  &  $-$0.06  &  $-$0.02 & B8  & Si\,{\sc ii} a little weak\\
14  & 0 55 33.921  & $-$37 43 11.63 &   19.72  & $-$0.26  & \nodata &   & H II region\\
15  & 0 55 31.716  & $-$37 43 23.95 &   21.00  & $-$0.12 & \nodata & B3  & noisy, Balmer lines strong\\
16  & 0 55 29.379  & $-$37 43 57.70 &   21.29  &  0.15  &   0.39 & F Ib ? & Balmer lines strong\\
17  & 0 55 28.182  & $-$37 44 23.41 &   19.78  & $-$0.01   &  0.10 & B3  & \\
18  & 0 55 34.875  & $-$37 44 48.15 &   19.99  & $-$0.03     & 0.01 & B8  & \\
19  & 0 55 38.936  & $-$37 44 57.33 &   20.75  &  0.12 & \nodata & A8--F0  & \\
FIELD B & & & & & & & \\
1   & 0 55 00.217  & $-$37 37 03.26 &   20.54  &  0.52 & \nodata & late-type  & \\
2   & 0 55 01.823  & $-$37 37 24.07 &   20.06  & $-$0.08  &   0.05 & B8  & He\,{\sc i} 4713 weak\\
3   & 0 54 46.351  & $-$37 37 51.35 &   20.03  & $-$0.08   &   0.06 & B3--B5  & He\,{\sc i} 4121,4143 missing\\
4   & 0 54 51.799  & $-$37 38 13.93 &   20.09  & $-$0.05 & \nodata & B1  & nebular contamination\\
5   & 0 54 44.166  & $-$37 38 31.53 &   19.83  & $-$0.04  &  $-$0.03 & B8  & \\
6   & 0 54 45.185  & $-$37 38 47.96 &   18.97  & $-$0.14 & \nodata &   & H II region\\
7   & 0 54 51.950  & $-$37 39 23.85 &   20.45  & $-$0.04 & \nodata & B2  & somewhat noisy\\
8   & 0 55 00.527  & $-$37 39 40.54 &   19.83  & $-$0.01 & \nodata & A0  & nebular contamination?\\
9   & 0 55 03.270  & $-$37 39 56.00 &   19.97  &  0.16  & 0.33 & A8  & Balmer lines somewhat strong\\
10  & 0 54 47.072  & $-$37 40 25.63 &   19.45  &  0.05 & \nodata & A2--A3  & \\
11  & 0 54 47.310  & $-$37 40 42.28 &   19.95  & 0.00 & \nodata & A5  & \\
12  & 0 54 56.083  & $-$37 41 04.02 &   19.30  & $-$0.18 & 0.00 & B0.5  & \\
13  & 0 54 56.754  & $-$37 41 33.79 &   18.64  & $-$0.11  & \nodata & B3  & Balmer lines filled\\
14  & 0 55 01.465  & $-$37 41 46.80 &   19.86  &  0.04   & \nodata & A0  & nebular contamination\\
15  & 0 55 03.464  & $-$37 42 26.18 &   19.71  & $-$0.06 & \nodata & B1.5--B2  & nebular contamination\\
16  & 0 54 44.645  & $-$37 42 39.13 &   19.02  & $-$0.11 & \nodata &   & emission line object, WN11\\
17  & 0 55 04.077  & $-$37 42 53.99 &   19.37  & $-$0.10 & \nodata & B0.5  & Balmer lines filled, P Cyg profiles at\\
& & & & & & & He\,{\sc i} 4471,4713?\\
18  & 0 54 58.692  & $-$37 43 21.06 &   20.35  & $-$0.17 & \nodata & B2--B3   & \\
19  & 0 54 57.705  & $-$37 43 38.65 &   19.93  & $-$0.05 & \nodata & B8--B9  & \\
FIELD C & & & & & & & \\
1   & 0 54 27.873  & $-$37 32 30.71 &   19.14  &  0.07 & 0.17 & A1  & Balmer lines strong\\
2   & 0 54 41.680  & $-$37 33 45.40 &   21.28  &  0.09 & 0.38 & A4  & Balmer lines strong\\
3   & 0 54 36.242  & $-$37 33 51.58 &   19.88  & 0.03   & 0.14 & B8--B9  & Si\,{\sc ii} weak, nebular contamination\\
4   & 0 54 41.759  & $-$37 34 36.85 &   21.31  & $-$0.03  &  0.13 &   & composite?\\
5   & 0 54 31.834  & $-$37 34 32.86 &   20.66  & $-$0.12  & 0.40 & B2  & metal poor\\
6   & 0 54 33.190  & $-$37 34 55.72 &   19.91  &  0.05  & 0.27 & A0  & metal depleted?\\
7   & 0 54 39.558  & $-$37 35 39.28 &   20.35  & 0.00  &  0.15 & A0  & Balmer lines strong\\
8   & 0 54 23.455  & $-$37 35 07.01 &   20.05  & $-$0.05  &  0.08 & B9  & metal poor? Balmer lines strong\\
9   & 0 54 21.294  & $-$37 35 17.93 &   20.25  &  0.02  &  0.19 & B9  & Balmer lines strong\\
10  & 0 54 32.431  & $-$37 36 20.43 &   21.03  &  0.27  & 0.52 &   & composite?\\
11  & 0 54 21.331  & $-$37 35 56.24 &   19.66  & $-$0.01   & 0.35 & B4  & \\
12  & 0 54 18.050  & $-$37 36 09.71 &   20.16  &  0.02  &  0.19 & B9--A0  & Mg\,{\sc ii} weak\\
13  & 0 54 31.605  & $-$37 37 44.29 &   19.75  & $-$0.16 & \nodata & B2  & Balmer lines broad, metal poor\\
14  & 0 54 26.881  & $-$37 37 34.41 &   20.26  & $-$0.18  & $-$0.07 & B1  & \\
15  & 0 54 22.197  & $-$37 37 44.98 &   20.99  & $-$0.21 & \nodata & B2  & for Gal. abundance, B1--B1.5 if metal-poor\\
16  & 0 54 21.938  & $-$37 38 12.77 &   18.04  &  0.05   & 0.19 & B9  & Balmer lines filled\\
17  & 0 54 22.528  & $-$37 38 25.50 &   20.97  &  0.13  &  0.31 &   & composite?\\
18  & 0 54 25.855  & $-$37 39 19.06 &   20.03  & $-$0.25  &  $-$0.14 & O9  & \\
19  & 0 54 24.039  & $-$37 39 17.13 &   21.42  & $-$0.21 & \nodata & B2.5  & \\
FIELD D & & & & & & & \\
1   & 0 54 34.625  & $-$37 39 19.70 &   21.00  & 0.05 & \nodata &   & composite?\\
2   & 0 54 26.406  & $-$37 39 52.05 &   19.90  &  0.03 & \nodata & A1  & Balmer lines somewhat strong\\
3   & 0 54 31.848  & $-$37 40 00.89 &   21.50  &  0.03  & 0.12 &   & composite?\\
4   & 0 54 23.153  & $-$37 40 46.30 &   19.85  & $-$0.12 & \nodata &   & H II region\\
5   & 0 54 25.597  & $-$37 41 03.44 &   19.97  & $-$0.05 & \nodata & B4  & for Gal. abundance; B3 if metal-poor\\
6   & 0 54 30.934  & $-$37 41 06.53 &   20.47  & 0.00 & \nodata & A1  & composite?\\
7   & 0 54 21.956  & $-$37 41 45.67 &   20.04  &  0.14 & \nodata & A4  & \\
8   & 0 54 31.235  & $-$37 41 55.96 &   19.52  & $-$0.12 & \nodata & B1.5  & \\
9   & 0 54 29.782  & $-$37 42 19.24 &   21.38  & $-$0.23 & \nodata & B1.5  & noisy\\
10  & 0 54 32.251  & $-$37 42 49.26 &   19.81  &  0.11  &  0.23 & A4  & \\
11  & 0 54 26.794  & $-$37 43 27.02 &   20.64  & $-$0.24 & \nodata &   & H II region\\
12  & 0 54 32.423  & $-$37 43 37.46 &   18.65  &  0.17  & 0.29 & A2  & Balmer lines filled\\
13  & 0 54 26.105  & $-$37 43 56.84 &   18.98  &  0.03  &  0.14 & A0  & \\
14  & 0 54 36.371  & $-$37 44 02.97 &   19.09  & $-$0.14   & $-$0.17 & B3  & metal poor, Balmer lines broad\\
15  & 0 54 40.048  & $-$37 44 23.12 &   20.61  & $-$0.11  &  $-$0.08 & B8  & Si\,{\sc ii} weak: metal deficiency?\\
16  & 0 54 32.892  & $-$37 45 02.51 &   20.92  &  0.06  &  0.18 & A2--A3  & \\
17  & 0 54 32.693  & $-$37 45 21.46 &   19.54  &  0.02   & 0.13 & B9--A0  & \\
18  & 0 54 21.804  & $-$37 46 04.60 &   19.83  & $-$0.01  &  0.08 & B9  & metal poor\\
\enddata

\end{deluxetable}

\clearpage
\begin{deluxetable}{lrr}
\tabletypesize{\scriptsize}
\tablecaption{A-8 and D-13: basic properties and stellar parameters \label{tab3}}
\tablewidth{0pt}
\tablehead{
\colhead{Star} & \colhead{A--8}   & \colhead{D--13}
}
\startdata
Spectral Type                &     B9--A0\,Ia       &      A0\,Ia\\
RA\,(J2000)                  &     0 55 41.619      &      0 54 26.105\\
DEC\,(J2000)                 & $-$37 40 58.61       &  $-$37 43 56.84\\
$v_{\rm rad}$ (km\,s$^{-1}$) & $+$122               &  $+$189\\[2mm]
{\em Atmospheric:}\\
$T_{\rm eff}$ (K)            & 10000\,$\pm$\,300    &  9500\,$\pm$\,300\\
$\log g$ (cgs)               & 1.60\,$\pm$\,0.15    &  1.35\,$\pm$\,0.15\\
$\xi$ (km\,s$^{-1}$)         & 8                    &  8\\
$y$ (by number)              & 0.09                 &  0.12\\
$[$M/H$]$ (dex)              & $-$0.7\,$\pm$\,0.2   &  $-$0.3\,$\pm$\,0.2\\[2mm]
{\em Photometric:}\\
$V$                          & 19.44\,$\pm$\,0.03   &  18.98\,$\pm$\,0.03\\
$B-V$                        & $-$0.01\,$\pm$\,0.03 &  0.03\,$\pm$\,0.03\\
$E(B-V)$                     & 0.02\,$\pm$\,0.03    &  0.03\,$\pm$\,0.03\\
$M_{\rm V}$                  & $-$7.15\,$\pm$\,0.12 &  $-$7.64\,$\pm$\,0.12\\
$B.C.^{\rm a}$               & $-$0.47              &  $-$0.41\\
$M_{\rm bol}$                & $-$7.62\,$\pm$\,0.12 &  $-$8.05\,$\pm$\,0.12\\[2mm]
{\em Physical:}\\
$\log L/$L$_{\odot}$         & 4.90\,$\pm$\,0.05    & 5.08\,$\pm$\,0.05\\
$R/$R$_{\odot}$              & 94\,$\pm$\,8         & 128\,$\pm$\,11\\
$M^{\rm ZAMS}/M_{\odot}^{\;\;b}$ & 16\,$\pm$\,3         & 19\,$\pm$\,3\\
$M^{\rm spec}/M_{\odot}$   & 13\,$\pm$\,2         & 13\,$\pm$\,3\\
\enddata

\tablenotetext{a}{from Schmidt-Kaler (1982)}
\tablenotetext{b}{from comparison with stellar evolution tracks from
Meader \& Meynet (2001)}

\end{deluxetable}

\end{document}